\documentclass[preprint,superscriptaddress,
amsmath,amssymb,
aps,prb,
floatfix,
showpacs,
]{revtex4-1}

\usepackage[utf8]{inputenc}
\usepackage[T1]{fontenc}

\usepackage{graphicx}
\usepackage{xcolor}
\graphicspath{{./graphics/}}

\usepackage{dcolumn}	
\usepackage{bm}			
\usepackage{hyperref}	
\usepackage{color}
\usepackage{siunitx}

\usepackage[labelfont=bf]{caption}

\begin{document}

\title{Direct observation of handedness-dependent quasiparticle interference in the two enantiomers of topological chiral semimetal PdGa}
\author{Paolo Sessi} 
\affiliation{Max Planck Institute of Microstructure Physics, Halle 06120, Germany}
\author{Feng-Ren Fan} 
\affiliation{Max Planck Institute for Chemical Physics of Solids, Dresden 01187, Germany}
\author{Felix K{\"u}ster} 
\affiliation{Max Planck Institute of Microstructure Physics, Halle 06120, Germany}
\author{Kaustuv Manna} 
\affiliation{Max Planck Institute for Chemical Physics of Solids, Dresden 01187, Germany}
\author{Niels B.M. Schr{\"o}ter} 
\affiliation{Swiss Light Source, Paul Scherrer Institute, CH-5232 Villigen PSI, Switzerland}
\author{Jing-Rong Ji} 
\affiliation{Max Planck Institute of Microstructure Physics, Halle 06120, Germany}
\author{Samuel Stolz} 
\affiliation{EMPA, Swiss Federal Laboratories for Materials Science and Technology, 8600 D{\"u}bendorf, Switzerland}
\affiliation{Institute of Condensed Matter Physics, Station 3, EPFL, 1015 Lausanne, Switzerland}
\author{Jonas A. Krieger} 
\affiliation{Swiss Light Source, Paul Scherrer Institute, CH-5232 Villigen PSI, Switzerland}
\affiliation{Laboratory for Muon Spin Spectroscopy, Paul Scherrer Institute, CH-5232 Villigen PSI, Switzerland}
\affiliation{Laboratorium für Festk{\"o}rperphysik,  ETH Zurich, CH-8093 Zurich, Switzerland}
\author{Ding Pei} 
\affiliation{Clarendon Laboratory, Department of Physics, University of Oxford, Oxford OX1 3PU, United Kingdom}
\author{Timur K. Kim} 
\affiliation{Diamond Light Source, Didcot, OX110DE, United Kingdom}
\author{Pavel Dudin} 
\affiliation{Diamond Light Source, Didcot, OX110DE, United Kingdom}
\author{Cephise Cacho} 
\affiliation{Diamond Light Source, Didcot, OX110DE, United Kingdom}
\author{Roland Widmer} 
\affiliation{EMPA, Swiss Federal Laboratories for Materials Science and Technology, 8600 D{\"u}bendorf, Switzerland}
\author{Horst Borrmann} 
\affiliation{Max Planck Institute for Chemical Physics of Solids, Dresden 01187, Germany}
\author{Wujun Shi} 
\affiliation{School of Physical Science and Technology, ShanghaiTech University, 201203 Shanghai, China}
\author{Kai Chang} 
\affiliation{Max Planck Institute of Microstructure Physics, Halle 06120, Germany}
\author{Yan Sun} 
\affiliation{Max Planck Institute for Chemical Physics of Solids, Dresden 01187, Germany}
\author{Claudia Felser} 
\affiliation{Max Planck Institute for Chemical Physics of Solids, Dresden 01187, Germany}
\author{Stuart S.P. Parkin} 
\affiliation{Max Planck Institute of Microstructure Physics, Halle 06120, Germany}


\maketitle

\vspace{1cm}
{\bf It has recently been proposed that combining chirality with topological band theory may result in a totally new class of fermions \cite{PhysRevB.94.195205,PhysRevB.93.045113,PhysRevLett.119.206402,CWS2018}. These particles have distinct properties: they appear at high symmetry points of the reciprocal lattice, they are connected by helicoidal surface Fermi arcs spanning the entire Brillouin zone, and they are expected to exist over a large energy range \cite{PhysRevLett.119.206402,CWS2018,SBC2019,SPV2019}.  Additionally, they are expected to give rise to totally new effects forbidden in other topological classes \cite{DGM2017,PhysRevLett.119.107401,PhysRevLett.116.077201}. Understanding how these unconventional quasiparticles propagate and interact is crucial for exploiting their potential in innovative chirality-driven device architectures. These aspects necessarily rely on the detection of handedness-dependent effects in the two enantiomers and remain largely unexplored so far.  Here, we use scanning tunnelling microscopy to visualize the electronic properties of both enantiomers of the prototypical chiral topological semimetal PdGa at the atomic scale. We reveal that the surface-bulk connectivity goes beyond ensuring the existence of topological Fermi arcs,  but also determines how quasiparticles propagate and scatter at impurities, giving rise to chiral quantum interference patterns of opposite handedness and opposite spiralling direction for the two different enantiomers, a direct manifestation of the change of sign of their Chern number.  Additionally, we demonstrate that PdGa remains topologically non-trivial over a large energy range, experimentally detecting Fermi arcs in an energy window of more than 1.6 eV symmetrically centerd around the Fermi level. These results are rationalized in terms of the deep connection between chirality in real and reciprocal space in this class of materials, and they allow to identify PdGa as an ideal topological chiral semimetal.}

The discovery of symmetry-protected topological materials represents a milestone in condensed matter physics \cite{RevModPhys.82.3045, RevModPhys.83.1057}. They provide a paradigm twist to the topic of the band structure of solids, allowing for a classification of materials based on well-defined topological invariants that are calculated as global quantities from their bulk wave functions. On a fundamental level, the rise of topology in condensed matter has set a fertile ground for the realization, in table-top experiments, of concepts such as Majorana \cite{Mourik1003,Nadj-Perge602} and Weyl fermions \cite{Xu613,PhysRevX.5.031013}, which were first predicted but not realized in the context of high-energy physics. More recently, it has been suggested that condensed matters systems can host totally new fermions, resulting from band crossing protected by specific symmetries of one of the 230 space groups \cite{Bradlynaaf5037}. In this context, the concept of chirality occupies a primary role. Chiral structures are characterised by a well-defined handedness due to the lack of both mirror and inversion symmetries, resulting in two distinct  enantiomers. Their handedness can be manifested in several forms including non-collinear spin textures \cite{M915}, magnetochiral dichroism \cite{RR1997}, or unconventional superconductivity \cite{Carnicomeaar7969}. The additional existence of topologically non-trivial bands is expected to confer on chiral crystals unique physical properties which not only don't exist in conventional materials, but are also forbidden in other topological classes \cite{CWS2018}. These phenomena are directly linked to the Chern number, an integer used to classify the topological properties of the band structures in solids, and which is obtained by integrating the Berry curvature over a closed surface in momentum space.  Because of the pseudovector-character of the Berry phase, the Chern number reverts its sign under a mirror operation.  Far from being a purely mathematical concept, this property has far reaching implications and, in topologically non-trivial chiral crystals, is expected to result in radically new effects such as quantized circular photogalvanic effects \cite{DGM2017},  unusual phonon dynamics \cite{PhysRevLett.119.107401}, and gyrotropic magnetic effects \cite{PhysRevLett.116.077201}.

Angle resolved photoemission (ARPES) studies have provided strong evidence for chiral Fermi arcs in chiral crystals belonging to the space group $P2_13$, number 198 \cite{SBC2019,SPV2019}. Recent scanning tunnelling microscopy (STM) measurements also detected surface-orientation dependent states exhibiting chiral fermion characteristics in one CoSi enantiomer \cite{Yuaneaaw9485}. However, the experimental investigation of chirality-dependent phenomena, i.e. the emergence, observation, and manipulation of effects directly linked to the sign of the Chern number, remains largely unexplored so far. This calls for chiral topological semimetals for which both enantiomers can be selectively synthesised to control the sign of the topological charge while keeping all other material properties unchanged \cite{LSR2019}. Here, we use scanning tunnelling microscopy (STM) to visualise how these unconventional quasiparticles propagate and interact with defects in the two enantiomers of the prototypical chiral semimetal PdGa. Our results provide compelling experimental evidence of a new and distinct feature of this class of materials: handedness-dependent scattering. We directly detect this effect in two distinct ways: (i) the opposite chirality of the quantum coherent interference patterns of the surface Fermi arcs in the two enantiomers and, (ii) their opposite energy-dependent spiralling direction, i.e. clockwise and anticlockwise. By imaging the perturbation pattern developing around defects in the crystal lattice of the two enantiomers, our results provide a self-consistent experimental evidence of the deep connection between chirality in real and reciprocal space in this class of materials, with chiral multifold-fermion crossing in reciprocal space being protected by the chiral crystal structure in real space. Finally, by spectroscopically analysing unoccupied states, which are not accessible by conventional photoemission techniques, we demonstrate that PdGa remains topologically non-trivial over a very wide energy window, with Fermi arcs existing in an energy window of more than 1.6 eV symmetrically centerd around the Fermi level.  These observations, jointly with the large extension of the Fermi arcs over the surface Brillouin zone and the minimum number of topologically protected band crossing, set PdGa as an ideal topological conductor, making it a promising platform to access and utilize optical and transport phenomena dictated by its topology.

\begin{figure*}[h!]
        \renewcommand{\figurename}{Figure}
	\centering
	\includegraphics[width=.9\textwidth]{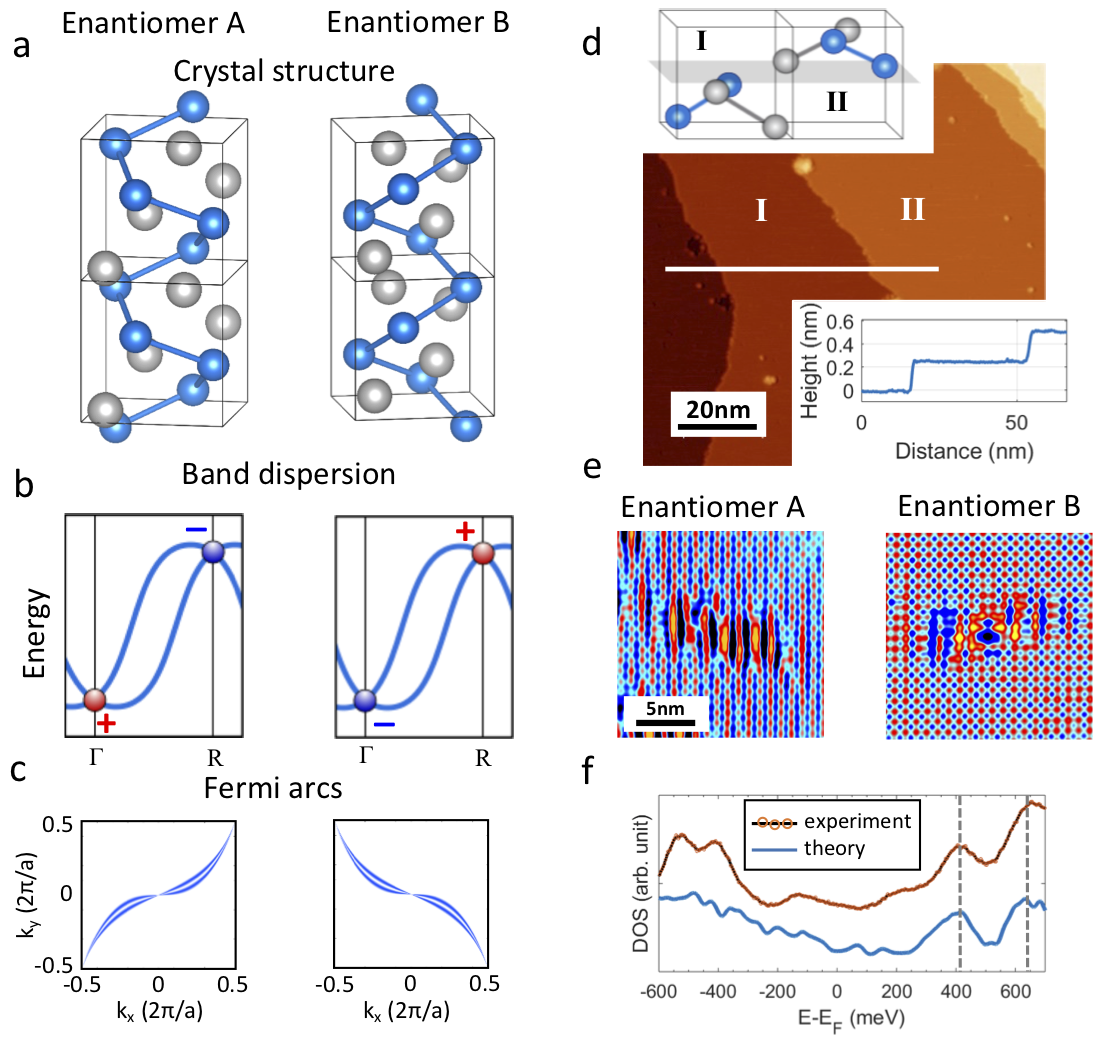}
	\caption{{\bf Connection between chirality in real and reciprocal space in PdGa}. {\bf a}  Crystal structure for both PdGa enantiomers. Gray and blue atoms correspond to Pd and Ga, respectively. The handedness can be distinguished considering the helix formed by Ga atoms; {\bf b}  Schematic line cut showing the symmetry-protected band crossing at  $\Gamma$ and R points, respectively. The Chern number associated to the nodes reverts its sign by mirror operation. {\bf c}  Fermi arcs developing on the surface for the two different enantiomers as a result of the surface-bulk connectivity. {\bf d}  Topographic overview of the PdGa (001) surface. The line profile evidence the minimum step height, corresponding to half unit cell. The two possible surface terminations, labelled I and II, are reported in the inset. {\bf e}  Atomically resolved images for crystals of opposite handedness. The perturbation developing around native defects can not be superimposed to its mirror image. {\bf f}  Comparison between experimental (red line) and theoretical (blue line) local density of states.}
	\label{Figure1}
\end{figure*}

PdGa belongs to the family of chiral crystals with a cubic B20 structure as illustrated in Fig. 1a. Grey and blue correspond to Pd and Ga atoms, respectively. The chirality can be distinguished by the handedness of the helix such as that formed by the Ga atoms, which rotates either clockwise or anticlockwise depending on the enantiomers. The structural chirality directly determines the electronic properties. As schematically illustrated in Fig. 1b, symmetry protected band crossings are visible at the $\Gamma$ and R high symmetry points in the bulk. In particular, a spin-3/2 fermion is realized near the $\Gamma$ point, while a double spin-1 fermion is realized at the R (see Supplementary Information, Section III for a detailed discussion). These crossings act as a source (red dot) or a sink (blue dot) of Berry curvature. They correspond to a Chern number of magnitude 4, i.e. the maximum Chern number achievable at a multifold node crossing, and reverse sign under a mirror operation \cite{schrter2019}. Note that, contrary to non-chiral topological semimetals, these crossings are well-separated in energy. Because of the surface-bulk correspondence characterizing topological materials, this scenario results in the emergence of topologically protected surface Fermi arcs emanating from momenta that match that of the surface projections of the bulk’s nodes. At the (001) surface, $\Gamma$ and R bulk points are maximally separated, being projected onto the $\overline{\Gamma}$ and $\overline{\textrm{M}}$ points, respectively. Consequently, Fermi arcs characterized by an extremely large extension appear, spanning the whole surface Brillouin zone as illustrated in Fig. 1c, with a dispersion which  changes sign under a mirror operation. 

Fig. 1d shows a topographical overview of the PdGa(001) surface (experimental details on samples preparation can be found in Supplementary Information, Section I). Large, atomically flat terraces with extremely low defect concentrations are visible, confirming the high quality of the crystals. The line profile analysis allows the identification of the smallest step height, which is 2.5 \AA. This value matches well with that of half the unit cell as illustrated in the inset, where adjacent terraces are labeled I and II, respectively.  Atomically resolved images for crystals of opposite handedness are displayed in Fig.1e. A square lattice is visible in both enantiomers with a periodicity $a =$  4.9 \AA \,matching the bulk lattice constant. A careful inspection of the perturbation developing around native defects reveals the presence of a strongly anisotropic pattern which can not be superimposed to its mirror image. As shown in the Supplementary Information (Section II) this behaviour is consistent with the bulk characterization of the two enantiomers, providing a direct real space signature of the structural bulk chirality. 

The local density of states (LDOS) has been experimentally inferred by scanning tunnelling spectroscopy (STS) measurements. Results are reported in Fig.\,1f. The minimum in the LDOS that is visible around the Fermi level highlights the semimetallic character of the compound. Even though STS strongly depends on how electronic states decay into the vacuum, with higher sensitivity for states located at the center of the surface Brillouin zone \cite{PhysRevLett.75.2960}, our results are in good agreement with the theoretically calculated LDOS, obtained by projecting the bulk band structure over the surface. A one-to-one matching is evident for all of the most prominent features visible in the spectrum, i.e. the peaks located at +400 and +650 meV with respect to the Fermi level (see grey dashed lines).  

\begin{figure*}
        \renewcommand{\figurename}{Figure}
	\centering
	\includegraphics[width=.9\textwidth]{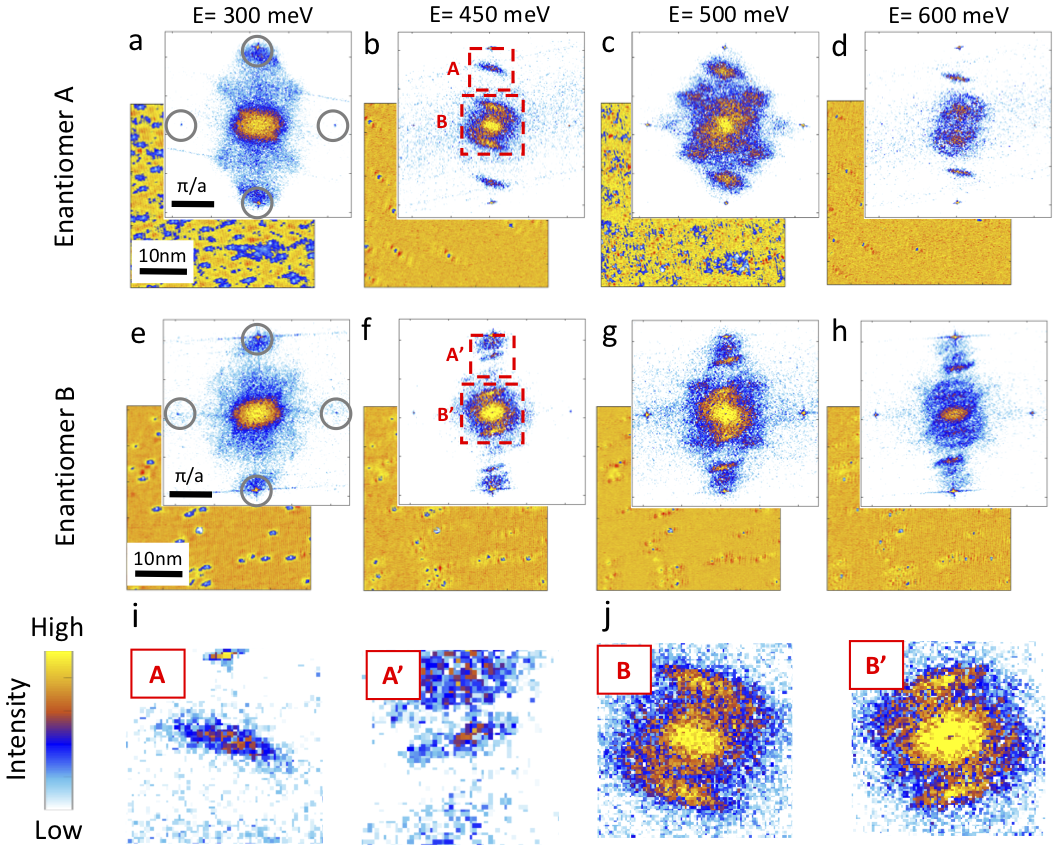}
	\caption{{\bf Quasiparticle interference of PdGa(001) enantiomers}. {\bf a-h} d$I$/d$U$ maps and relative Fourier transformations obtained on PdGa(001) with opposite bulk chirality at four representatives energies. Both long A(A') and short wavelength B(B') scattering vectors are visible, whose shape rapidly evolve with energy. Bragg spots (see grey circles) are representatives of the surface square lattice. As highlighted in panels {\bf i} and {\bf j}, the scattering vectors are chiral.}
	\label{Figure2}
\end{figure*}

To investigate how the bulk chirality impacts the electronic properties, we analyzed the standing wave patterns generated by coherent scattering of quasiparticles at defects. The resulting LDOS modulations are visualised by energy-resolved differential conductance (d$I$/d$U$) maps that were measured at 1.9 K in order to reach a large coherence length and an improved energy resolution. Fourier transforms (FT) of these data allow the quantitative analysis of this information into reciprocal space, making the visualisation of the scattering vectors $\bf{q}$ that connect the initial $\bf{k_i}$ and final $\bf{k_f}$ states on an isoenergy contour possible, i.e. $\bf{q}$ = $\bf{k_i}$ - $\bf{k}_f$. This technique, originally developed in the context of trivial surface states in noble metal surfaces \cite{CLE1993}, has recently been applied to investigate the unconventional electronic properties of different classes of topologically non-trivial materials \cite{RSP2009,PhysRevLett.103.266803,Okada1496,Inoue1184}. In contrast to conventional photoemission, this method allows access to both occupied and unoccupied states, thus providing a complete spectroscopic characterization of quasiparticles that sit close to the Fermi level, i.e. those dominating the transport properties.  This is particularly important in the present case. Photoemission studies showed that Fermi arcs are overlapping with topologically trivial bulk bands for occupied states, complicating their identification \cite{schrter2019}. On the other hand, theoretical calculations predict them to become strongly decoupled from bulk states at positive energies, a scenario favouring their experimental detection (see Supplementary Information, Section III).

Fig.2, a to h, summarises the results obtained at four representative energies on PdGa(001) single crystals of opposite chirality.  The Bragg spots of the square (001) surface lattice can clearly be recognised (highlighted by four grey circles). They are located at a distance $2\pi/a$ from the center, with $a =$  4.9\,\AA\,being the lattice constant. The FT-maps evidence a rich plethora of scattering vectors. Their length, and even more remarkably their shape, rapidly evolves with energy. No FT-pattern is visible for occupied states, a direct consequence of Fermi arcs overlapping with bulk states, generating a continuum of possible scattering vectors which washes out any Fermi-arc distinct feature (see Supplementary Information, Section III and IV).  For each energy, a one-to-one comparison of measurements taken on crystals of opposite handedness clearly reveals, despite different background contributions, that scattering events are chiral, i.e. FT-maps taken on the two enantiomers are related by a mirror operation. As highlighted in the zoomed FT-maps displayed in panel i and j, this is the case for both long and short scattering vectors, labelled A(A') and B(B'), respectively.

Measurements performed on terraces separated by half-integer unit cell provide the very same results (see Supplementary Information, Section V). This demonstrates that our findings are not related to possible terrace-dependent trivial surface states, but are an intrinsic property of the overall PdGa(001) surface termination. Additionally, Mn adatoms have also been dosed onto the surface. Despite this procedure increases the disorder, resulting in a significantly stronger background in FT-map, large $\bf{q}$ scattering vectors [labeled A(A') in Fig. 2] are still visible (see Supplementary Information, Section VI), confirming that the signatures visible in our FT-maps are of topological origin.

\begin{figure*}
        \renewcommand{\figurename}{Figure}
	\centering
	\includegraphics[width=\textwidth]{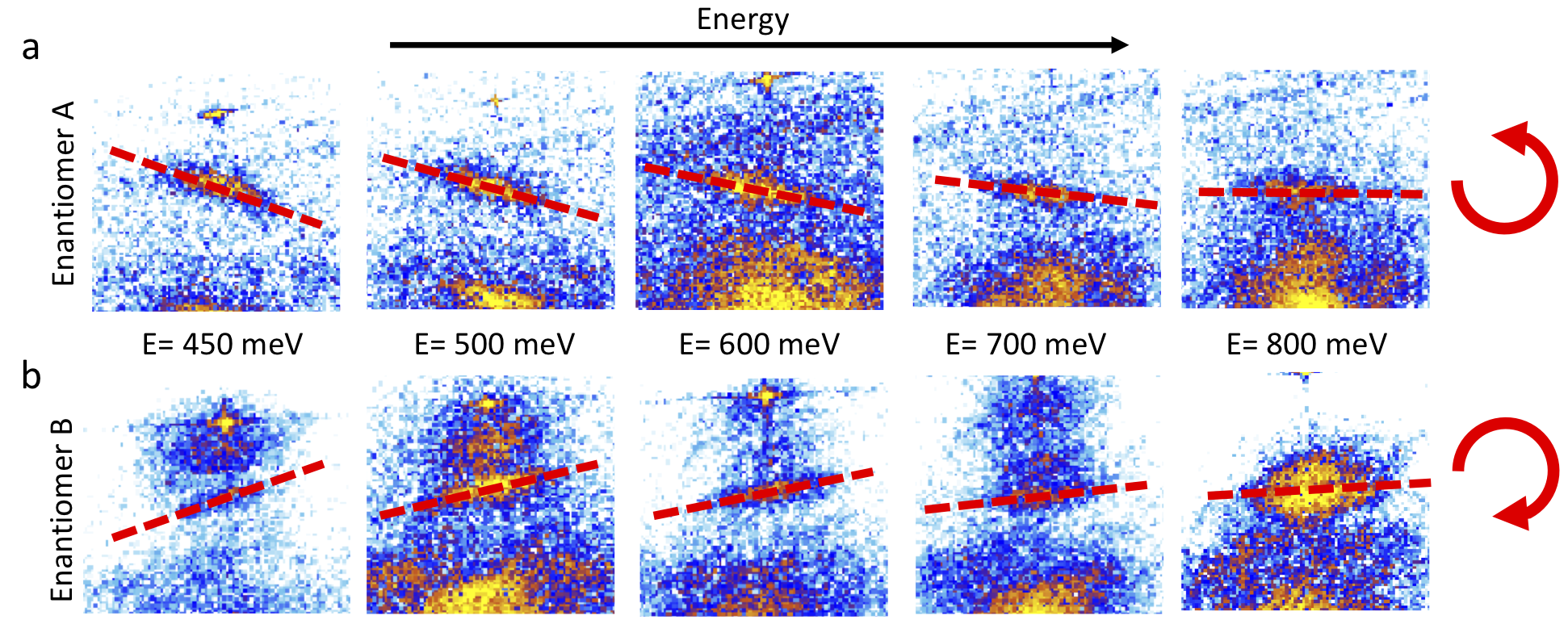}
	\caption{{\bf Spiralling direction of QPI patterns in the two enantiomers.} {\bf a} and {\bf b} report the energy evolution of the long scattering vector (A and A' in Fig. 2) for enantiomers A and B, respectively. By progressively increasing the energy, the scattering vector spirals in opposite directions,  i.e. anticlockwise and clockwise, in the two enantiomers. }
	\label{Figure4}
\end{figure*}

Furthermore, our measurements reveal opposite spiralling directions in the energy-dependent evolution of the QPI pattern for the two enantiomers, as illustrated in Fig. 3, where the long scattering vectors (A and A' in Fig. 2) are shown at progressively higher energies. A comparison between enantiomer A and B reveal their opposite rotational sense: anticlockwise vs. clockwise, respectively.

The emergence of chirality-dependent scattering between topological Fermi arcs is further supported by our theoretical analysis reported in Fig.\,4. Panel a reports the constant energy cut at E\,$ =+450$ meV  for both PdGa(001) enantiomers, showing long chiral Fermi arcs that are well decoupled from bulk states. As expected, only a mirror operation can convert one handedness of the Fermi arcs to the other. The FT-maps, calculated by including the spin-dependent scattering probabilities that take into account the influence of the relative spin orientations of the initial and final states, are displayed in Fig. 4b. The FT-maps, and thus the scattering events, are chiral in the two PdGa enantiomers and related by a mirror operation. The rich plethora of scattering vectors and their rapid evolution as a function of energy makes it difficult to establish a one-to-one correspondence for all features visible in the FT-d$I$/d$U$ maps. However, the large scattering vector [labeled A(A') in Fig. 2] is sufficiently decoupled from the typical background centerd around $\bf{q}$=0, which allows for a detailed theoretical analysis of its origin. This is illustrated in Fig. 4c-d, where different sections of the constant energy cuts are progressively included in our analysis. The resulting FT-patterns unambiguously prove that vectors A(A') are directly linked to scattering events between opposite surface Fermi arcs connected by good nesting vectors. 

\begin{figure*}
        \renewcommand{\figurename}{Figure}
	\centering
	\includegraphics[width=\textwidth]{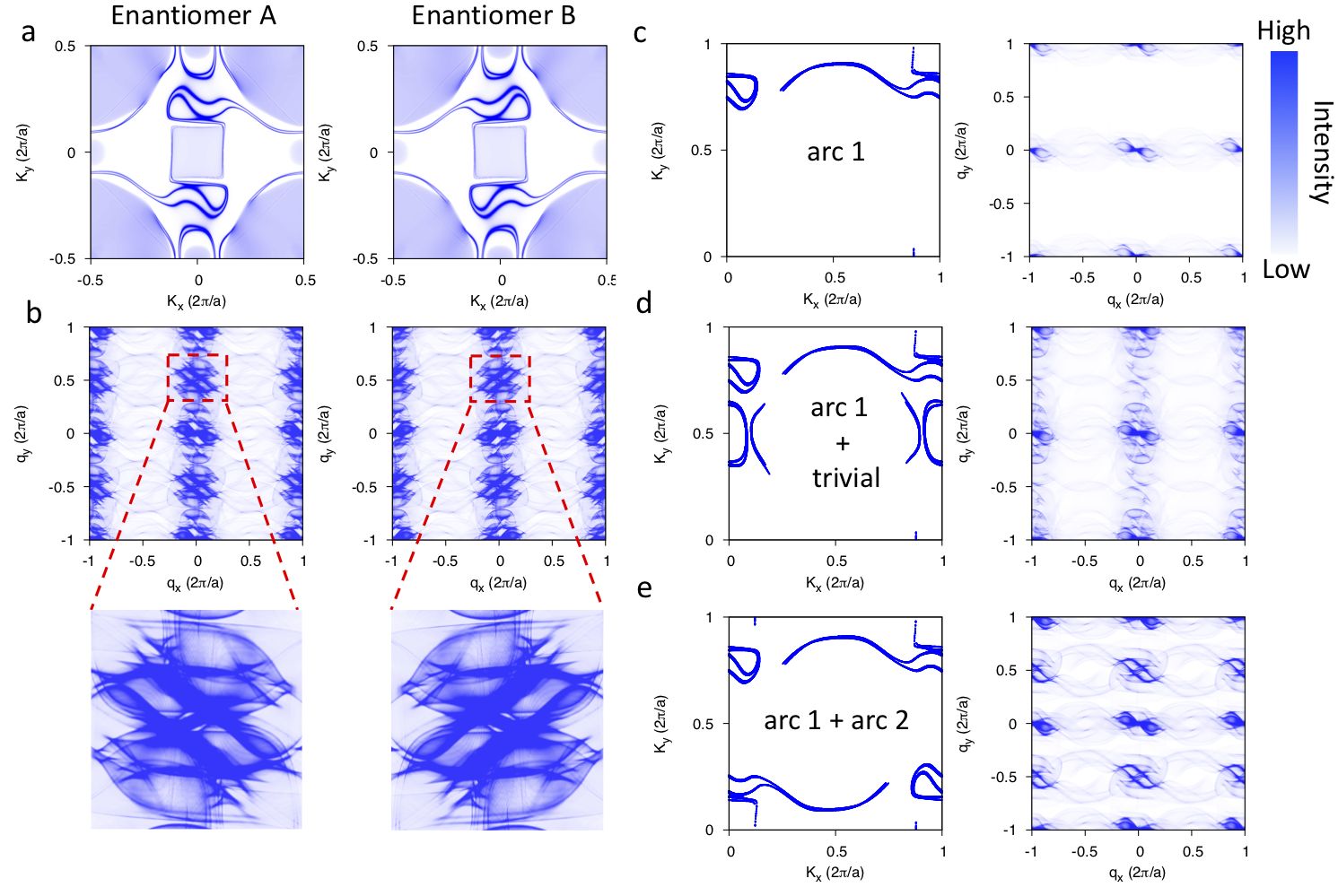}
	\caption{{\bf Fermi arcs and chiral quasiparticle interference}. {\bf a} Constant energy cuts at E = 450 meV for the two different PdGa(001) enantiomers. {\bf b} Theoretically calculated FT scattering maps. {\bf c-e} constant energy cuts and their relative scattering pattern by progressively including: {\bf c}, only one Fermi arc; {\bf d} one Fermi arc and trivial states; {\bf e} all Fermi arcs.}
	\label{Figure4}
\end{figure*}

\begin{figure*}[h!]
        \renewcommand{\figurename}{Figure}
	\centering
	\includegraphics[width=.95\textwidth]{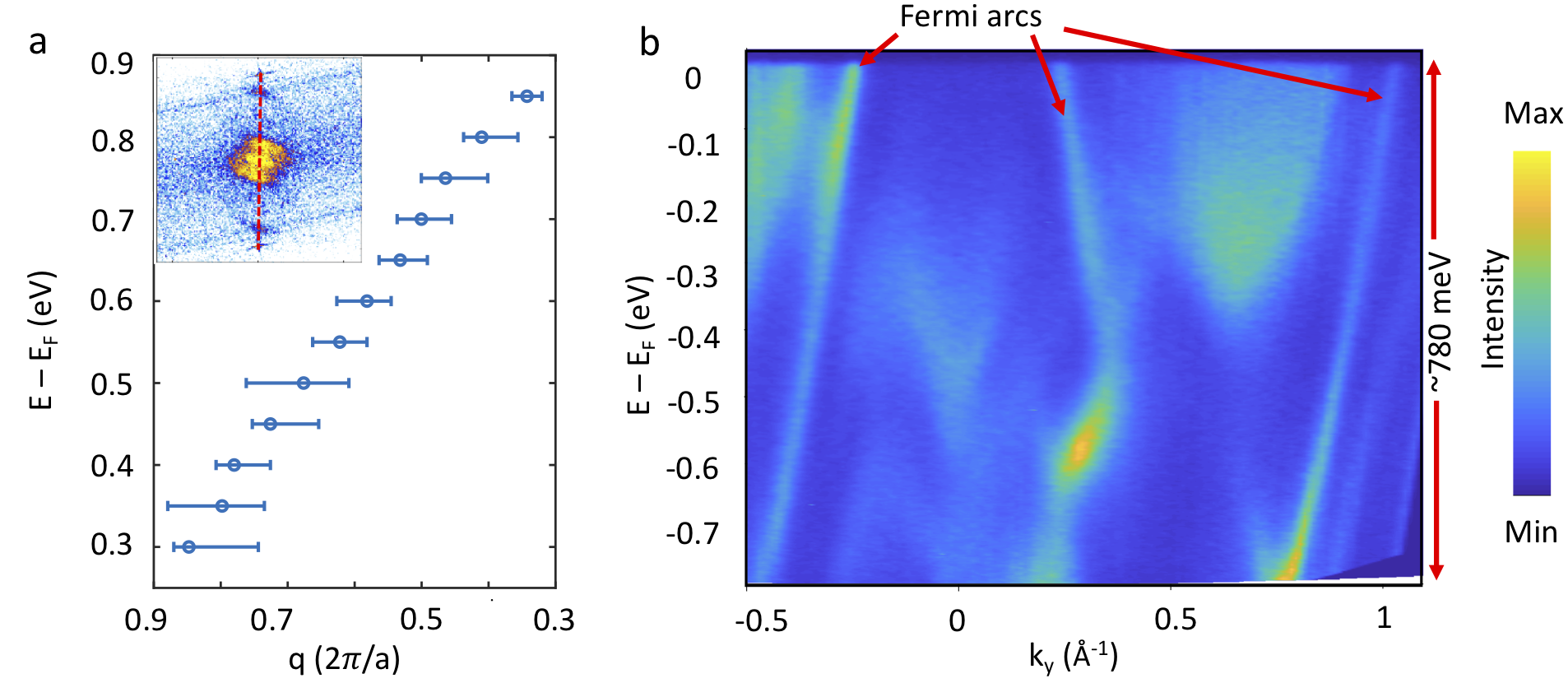}
	\caption{{\bf Energy window for Fermi arcs}. {\bf a} Energy dispersion of the scattering vector {\bf a} associated to scattering events between opposite Fermi arcs  (labeled A in Fig. 2) . The wavelength has been obtained, for all energies, by analysing the intensity profile taken along the red line passing through the center of the FT-d$I$/d$U$ map, as illustrated in the inset.  {\bf b} Energy dispersion of the Fermi arcs for occupied states as obtained by ARPES. The momentum direction was chosen along a path where the band bottom of the Fermi-arc becomes visible 
(see Supplementary Information for details). Measurements were performed with 60 eV photon energy and linear-horizontal polarization.}
	\label{Figure5}
\end{figure*}

The existence of quantum interference patterns originating from Fermi arcs allows their investigation as a function of energy. This is shown in Fig.\,5a, which shows the experimentally obtained energy-dependent length of the scattering vector A (see Supplementary Information, Section VII for a description of the analysis procedure).  These data reveal that Fermi arcs in PdGa are dominant up to 850 meV above the Fermi level. As mentioned above, not clear information can be obtained by QPI mapping for occupied states. As complementary information, we present ARPES data of PdGa crystals in panel b which shows that the band bottom of the Fermi arcs is located at approximately 780 meV below the Fermi level (see Supplementary Information, Section VIII, for a description of the measurement procedure). Overall,  this provides direct experimental evidence of one of the distinct features theoretically predicted for topological chiral crystals, i.e. the persistence of non-trivial Fermi arcs over a very large energy range, establishing an experimental record of more than 1.6 eV, almost symmetrically centred around the Fermi level.

Our work reveals the emergence of quantum interference in topological chiral crystals that depend on the crystal enantiomer. Consequently, in this class of materials, the surface-bulk correspondence not only guarantees the existence of topologically protected surface states, but also determines how they propagate and scatter. This phenomenon directly follows from the deep connection between chirality in real and reciprocal space, and is a direct manifestation of Chern numbers changing their sign by changing the handedness of the crystal structure. These findings, jointly with the the investigation of both PdGa enantiomers and the large extension of the Fermi arcs, constitute this material to be an ideal topological semimetal. This suggests that the topological response of PdGa can be accessed in transport and optical measurements \cite{chang2019}.

\vspace{1cm}
{\bf Acknowledgements:}
This work was financially supported by the Deutsche Forschungsgemeinschaft (DFG, German Research Foundation) – Project number 314790414.


\bibliography{bibliography}

\end{document}